# The High-resolution Fe K Spectrum of Cygnus X-3

Aswath Suryanarayanan[1] , Frits Paerels[1], and Maurice Leutenegger[2]
[1] Columbia Astrophysics Laboratory, Columbia University, 538 W. 120th St., New York, NY 10027, USA
[2] NASA/Goddard Space Flight Center, Code 662, Greenbelt, MD 20771, USA

## Abstract

We analyze features of the Fe K spectrum of the high-mass X-ray binary Cygnus X-3. The spectrum was obtained with the Chandra High Energy Transmission Grating Spectrometer in the third diffraction order. The increased energy resolution of the third order enables us to fully resolve the Fe XXV He$\alpha$ complex and the Fe XXVI Ly$\alpha$ lines. The emission-line spectrum shows the expected features of photoionization equilibrium, excited in the dense stellar wind of the companion star. We detect discrete emission from inner-shell transitions, in addition to absorption likely due to multiple unresolved transitions in lower ionization states. The emission-line intensity ratios observed in the range of the spectrum occupied by the Fe XXV $n = 1$–2 forbidden and intercombination lines suggest that there is a substantial contribution from resonantly scattered inner-shell emission from the Li- and Be-like ionization states. The Fe XXV forbidden and intercombination lines arise in the ionization zone closest to the compact object, and since they are not subject to radiative-transfer effects, we can use them in principle to constrain the radial velocity amplitude of the compact object. We infer that the results indicate a compact object mass of the order of the mass of the Wolf–Rayet companion star, but we note that the presence of resonantly scattered radiation from Li-like ions may complicate the interpretation of the He-like emission spectrum, and specifically of the radial velocity curve of the Fe XXV forbidden line.

*Unified Astronomy Thesaurus concepts:* High mass x-ray binary stars (733); Compact objects (288); High resolution spectroscopy (2096); Black holes (162)

## 1. Introduction

Cygnus X-3 (Cyg X-3) is a high-mass X-ray binary with a short orbital period of 4.8 hr (Liu et al. 2006). Cyg X-3's companion is a WN4-7 Wolf–Rayet (W-R) star (Koljonen & Maccarone 2017). This WN4-7 W-R star produces a massive stellar wind. Past literature on Cyg X-3 has estimated the stellar wind to have a velocity of the order of 1000 km s$^{-1}$. This wind is heavily ionized by X-ray radiation from the compact object. A notable feature of Cyg X-3 is the ambiguity surrounding the exact nature of the compact object, as existing limits on the mass of the compact object do not rule out either possibility—neutron star or black hole.

The 1.5–10 keV spectrum of Cyg X-3 observed with the Chandra High Energy Transmission Grating Spectrometer (HETGS) exhibits a detailed discrete emission spectrum (Paerels et al. 2000). Distinct features of the spectrum, like the intensity ratio of the He-like $n = 1$–2 emission lines, are consistent with the properties of cool, optically thin photoionized gas in photoionization equilibrium (Kawashima & Kitamoto 1996; Liedahl & Paerels 1996; Paerels et al. 2000). Most recently, Kallman et al. (2019) have analyzed the full 1–7 keV spectrum observed with HETGS in terms of a detailed spectroscopic model for X-ray photoionized gas.

The spectrum shows a net redshift of approximately 800 km s$^{-1}$ (Paerels et al. 2000; 550 km s$^{-1}$ according to Stark & Saia 2003), establishing that the ionization in the wind is asymmetric. These features of the wind (and its outflowing nature) are supported by the existence of P Cygni profiles (with blueshifted absorption) in multiple emission lines. Vilhu et al. (2009) suggest that the modulation of the broadband X-ray light curves from Cyg X-3, if any, can be attributed to electron scattering in the asymmetric wind.

In order to correctly measure the Doppler modulation of the compact object with respect to the W-R star, we need to analyze emission from gas located as close as possible to the compact object/X-ray continuum source. We can approximately identify the distribution of the emission lines with radial distance from the compact object using the ionization parameter. The degree of photoionization is governed by the density of the gas through the ionization parameter $\xi$, which is proportional to the ionizing flux and inversely proportional to the particle density, $\xi = L/(nr^2)$ (Tarter et al. 1969). With an ionizing source placed off-center in an outflowing wind, one generally expects an ionization structure centered on the X-ray source, with the ionization parameter decreasing away from the X-ray source (Hatchett & McCray 1977). The ionization distribution will be asymmetric with respect to the X-ray source, with a shape depending on the density distribution. The region of highest ionization, however, for most reasonable wind models, is compact and approximately spherical, centered on the X-ray source.

This suggests that potentially the best way to trace the radial velocity of the compact objects is to measure the orbital modulation of the radial velocity of the Fe XXV and Fe XXVI lines. Lower-charge ions may reside far away from the compact object, and their radial velocity curve cannot be used to measure the mass ratio unless the location of the ions with respect to the center of mass of the binary is somehow known. We choose to focus on the Fe XXV spectrum, since the optical depth of the wind in resonance lines, such as Fe XXVI Ly$\alpha$, may be nonzero and radiative-transfer effects may bias the emission-line profiles. The He-like Fe ion has strong $n = 1$–2 forbidden and intercombination transitions, which are not expected to be subject to this effect. We therefore concentrate on the Fe XXV spectrum.

However, without better resolution, the Doppler shifts in the Fe XXV complex cannot be separated uniquely for the different components of the emission-line complex: the forbidden (*z*), intercombination (*x*, *y*), and resonance (*w*) lines of Fe XXV. In first order, the full range of the Fe XXV $n = 1$–2 spectrum spans only 1.5 resolution elements in the spectrum registered with the HETG in the HETGS. This is also discussed in Vilhu et al. (2009), who do not work with Fe XXV due to the possibility of (unresolvable) selective absorption in the resonance line as a function of orbital phase.

While the third-grating-order spectra have lower intensity, the Cyg X-3 spectrum is bright enough to discern the individual components of the Fe XXV and Fe XXVI complexes. In this paper, we follow up results from past literature with an analysis of the third-order HETGS/HETG Fe K spectrum. Using the third order enables us to analyze features of the Fe spectrum with three times the resolving power of the first order and to separate the forbidden and intercombination lines from the resonance line. At 6700 eV ($\lambda = 1.850$ Å), the energy of the Fe XXV $n = 1$–2 resonance line, the resolving power in third order is $E/\Delta E = 460$, and the full range of the four $n = 1$–2 lines spans 4.5 resolution elements. We note that the second-order HETG spectrum has efficiency comparable to the third order, but it would only resolve the complex into three resolution elements, and we therefore focused on the third-order spectrum.

In the following, we will first discuss the general nature of the Fe K spectrum in Cyg X-3. The $m = 3$ HETGS spectrum is currently the only fully resolved Fe K spectrum in photoionization equilibrium from an astrophysical source, and it serves as an important template for the Fe K spectra of radiation-driven emission-line sources in general, such as we may expect from accreting sources of all varieties, including active galactic nuclei. We then turn to a detailed analysis of the optically thin emission lines, deriving constraints on the radial velocity curve for these lines. We conclude by interpreting the radial velocity constraints in terms of a lower limit on the mass of the compact object, in terms of the mass of the W-R companion. The most likely limits place the mass of the compact object at $M > 7.2 M_\odot$, arguing for a black hole.

## 2. Observation and Data Reduction

For our analysis, we extracted data from the Chandra X-ray Observatory archive. In particular, we worked with one Continuous Clocking (CC)–mode data set and one Timed Exposure (TE)–mode data set from a high state of Cyg X-3. In the CC mode of the ACIS/S detector, the CCDs are read out continuously in the cross-dispersion direction. These data therefore lack any information on the cross-dispersion intensity distribution of the spectrum, this information having been traded for increased time resolution. For very bright sources (such as Cyg X-3), in which the background is negligible, this does not produce difficulties in interpreting the spectrum; in fact, the rapid readout has the beneficial effect of mitigating complications due to pileup (see below). In the TE mode, the full spectral image is read out. For bright sources, the cross-dispersion integrated spectrum recorded in TE mode can be added to the CC-mode spectrum without difficulty.

Table 1 contains ObsID and exposure time details of the data sets. These data sets comprise about 75% of all counts obtained with HETGS on Cyg X-3 to date. Both the data sets have been used extensively by Vilhu et al. (2009) and Kallman et al. (2019). After downloading the data sets from the archive, we reprocessed the data sets (using the CIAO thread `chandra_repro`; we used CIAO 4.11 with CALDB 4.8.5 throughout). We then barycenter-corrected the reprocessed files in order to account for effects due to the Earth's orbit around the Sun and Chandra's orbit around Earth. Following this, we created good time intervals and aligned them if required (we aligned the ObsID 6601 data). Finally, we divided the spectrum into phase bins and analyzed the third-order phase-binned data sets using custom-written Python code. We combined the grating spectra from the two ObsIDs before analysis and divided the spectrum into two phase bins using the CIAO command `combine_grating_spectra`.

**Table 1**
Observations

| ObsID | Exp. Mode | Date | Net Exp. (ks) |
|---|---|---|---|
| 6601 | TE | 2006-01-26 | 49.56 |
| 7268 | CC | 2006-01-25 | 69.86 |

The complete imaging TE-mode data are more vulnerable to pileup associated with the finite spatial and temporal resolution of the CCD detectors than the CC-mode data. A general description of the effect can be found in the Chandra Proposers' Guide.[3]

Pileup is defined as the coincidence of two or more photons per CCD readout time, or frame time, within an event-detection cell. The detector will be unable to temporally resolve two or more photons, resulting in a charge deposition ("pulse height") that is roughly the sum of the pulse heights of the individual photon events. Hence, in the presence of pileup, the event-detection rate will be lower and the observed spectrum will be distorted toward higher energies. As a result, the observed count rate in very bright areas of the spectrum will be lower than the incident count rate.

According to Kallman et al. (2019), the high-energy grating (HEG) arm of the ObsID 6601 HETG spectrum is relatively unaffected by pileup. Since the Fe K lines are located above 6 keV, we use the HEG arm for all our data analyses. Regardless, we performed our own analysis of the spectrum for pileup, in order to verify the results obtained by Corrales & Paerels (2015). We utilized data from the brightest region of the spectra (3–5 keV) from both observations (see Figure 1). Note that the *third*-order Fe K photons (6600–6700 eV) will be superimposed on the first-order 2200–2333 eV energy range. We can determine whether ObsID 6601 is significantly affected by pileup by analyzing anomalies in the following ratio for every bin: $N_{6601}/N_{7268}$, where $N$ is the number of counts in that bin, in ObsIDs 6601 and 7268, respectively.

Figure 1 contains the results of this analysis for photon energies 2.9–6.7 keV, using the first-order HEG spectra. Since the observed count rate in very bright areas would be lower than the incident count rate for ObsID 6601 (this is not the case for ObsID 7268), a sustained statistically significant negative deviation from the mean ratio in a very bright region would be evidence of pileup. The ratio ranges from approximately $N(\text{TE})/N(\text{CC}) \approx 0.6$ at long wavelengths to about 0.7 at the shortest wavelengths, $\lambda = 1.5$–2.0 Å, which contains the Fe XXV spectrum. At the highest count rates in the TE-mode data, the emission lines between 3.0 and 3.2 Å (Ca XX, Ca XIX $n = 1$–2), the ratio is higher than average, implying that pileup is not important in the TE-mode spectrum. A quick analytical estimate confirms this: in the HETG m = +1-order TE-mode spectrum, at 2.8 Å, we have approximately 700 counts/(1 pixel bin), or 0.056 counts per CC readout per bin. At those low count rates, pileup is not significant. This analysis refers to first-spectral-order counts only, whereas it is the total (all orders) count per pixel that matters for pileup. The second-order count at the first-order 2.8 Å position is negligible, however: the first-order count at 1.4 Å, multiplied by the ratio of effective areas in m = +2 and m = +1 at 1.4 Å, amounts to approximately $1.7 \times 10^{-5}$ counts per bin per CCD readout. The third-order contribution from 0.93 Å is smaller still.

---
[3] https://cxc.cfa.harvard.edu/proposer/POG/html/chap6.html#tth_sEc6.16

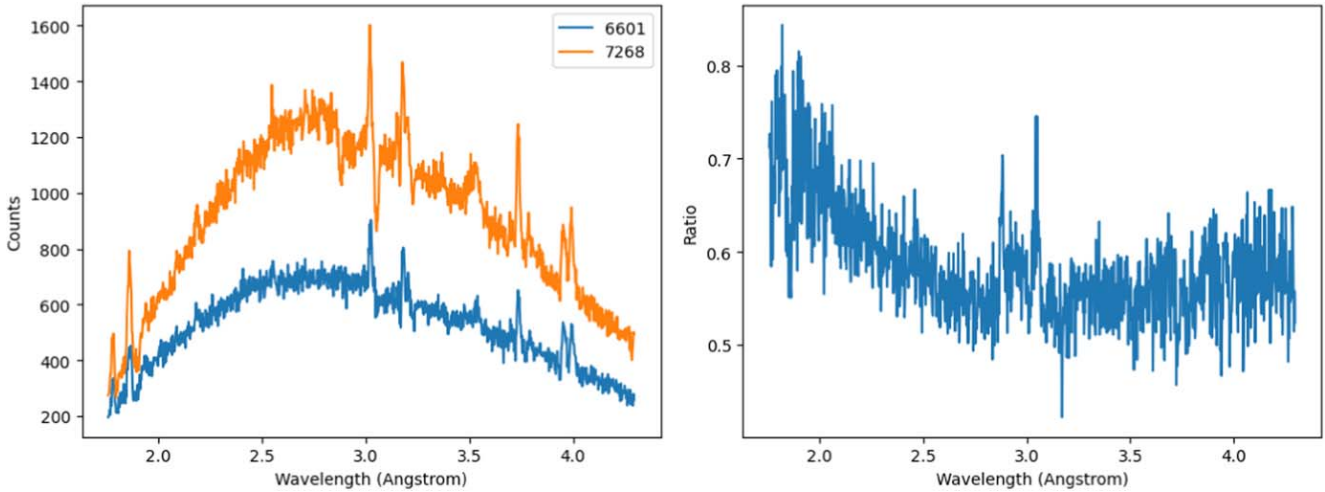

**Figure 1.** Left: spectra of ObsID 7268 (orange, CC mode) and ObsID 6601 (blue, TE mode). Right: graph of $N_{6601}/N_{7268}$.

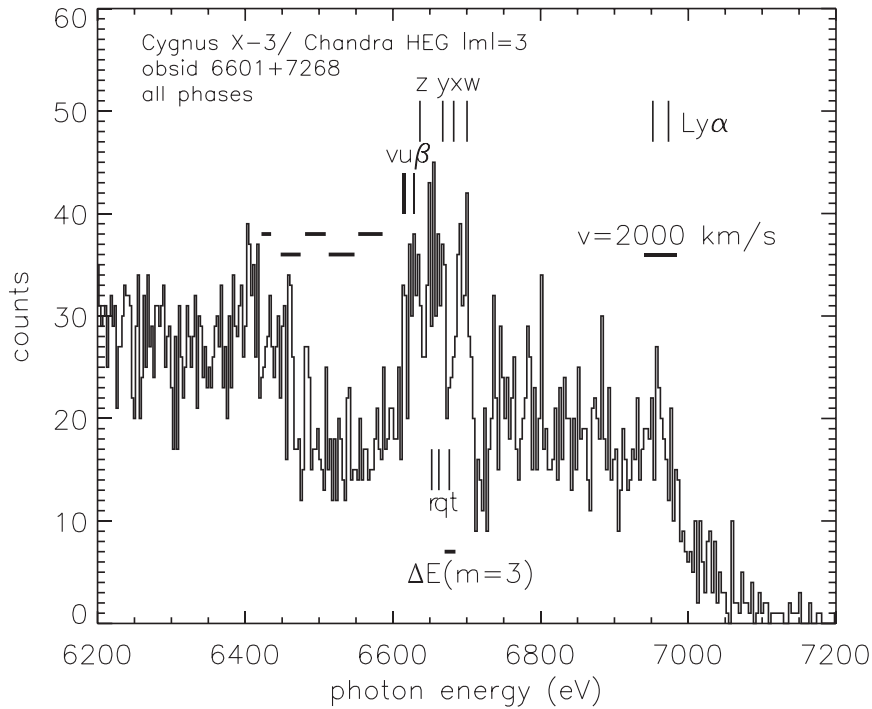

**Figure 2.** Chandra HETGS third-order spectrum of Cyg X-3. The spectrum has been binned in $8.33 \times 10^{-4}$ Å bins or $\Delta E \approx 3$ eV at 6700 eV. The nominal energies of a number of transitions have been indicated: the $n = 1$–2 transitions $w$, $x$, $y$, $z$ in He-like Fe, the $n = 1$–2 inner-shell transitions $u$, $v$, $r$, $q$ in Li-like Fe, and the $n = 1$–2 inner-shell transition $\beta$ in Be-like Fe. The spectrometer resolution in third order, $\Delta E \approx 15$ eV at 6700 eV, has been indicated. The horizontal bars in the range 6400–6600 eV indicate the range of $n = 1$–2 inner-shell transitions in the lower-charge-state members of the Fe L series, or B-like Fe through F-like Fe. The progression is from B-like Fe near 6600 eV to F-like Fe near 6400 eV. Energies were taken from Decaux et al. (1997). The background is mainly determined by the (nonimaging) CC-mode data and is dominated in this spectral range by the zero-order image of the interstellar dust-scattering halo. After passing through the CCD pulse height filter, that leaves a flux level of 1–2 counts per bin, the level approximately observed above 7100 eV.

## 3. The Phase-averaged Fe K Spectrum

In Figure 2, we show the 6200–7200 eV range of the third-order HETGS/HEG spectrum, with both ObsIDs combined, positive and negative orders added, and binned in $8.3 \times 10^{-4}$ Å bins ($\approx 3.0$ eV at 6700 eV). The summed spectrum covers

**Table 2**
Transitions $n = 1–2$ in the Range 6600–6700 eV

| Symbol | Energy[a] (eV) | Ion | Transition | $f$[b] | $y$[c] | Type |
|---|---|---|---|---|---|---|
| $v$ | 6613.31 | Li-like | $1s^2 2s\ ^2S_{1/2} - 1s2s2p\ ^4P_{1/2}$ | $2.62 \times 10^{-3}$ | 0.997 | IE |
| $u$ | 6616.629 | Li-like | $1s^2 2s\ ^2S_{1/2} - 1s2s2p\ ^4P_{3/2}$ | $1.63 \times 10^{-2}$ | 0.975 | IE |
| $\beta$ | 6628.804 | Be-like | $1s^2 2s^2\ ^1S_0 - 1s2s^2 2p\ ^1P_1$ | $7.14 \times 10^{-1}$ | 0.780 | IE |
| $z$ | 6636.78 | He-like | $1s^2\ ^1S_0 - 1s2s\ ^3S_1$ | $3.27 \times 10^{-7}$ | 1.00 | Forbidden |
| $r$ | 6652.826 | Li-like | $1s^2 2s\ ^2S_{1/2} - 1s2s2p\ ^2P_{1/2}$ | $1.57 \times 10^{-1}$ | 0.876 | IE |
| $q$ | 6662.240 | Li-like | $1s^2 2s\ ^2S_{1/2} - 1s2s2p\ ^2P_{3/2}$ | $4.90 \times 10^{-1}$ | 1.00 | IE |
| $y$ | 6667.671 | He-like | $1s^2\ ^1S_0 - 1s2p\ ^3P_1$ | $6.77 \times 10^{-2}$ | 1.00 | Intercombination |
| $t$ | 6676.202 | Li-like | $1s^2 2s\ ^2S_{1/2} - 1s2s2p\ ^2P_{1/2}$ | $9.63 \times 10^{-2}$ | 0.733 | IE |
| $x$ | 6682.67 | He-like | $1s^2\ ^1S_0 - 1s2p\ ^3P_2$ | $2.04 \times 10^{-5}$ | 1.00 | Intercombination |
| $w$ | 6700.549 | He-like | $1s^2\ ^1S_0 - 1s2p\ ^1P_1$ | $7.11 \times 10^{-1}$ | 1.00 | Resonance |

**Notes.**
[a] Measured energy, from Rudolph et al. (2013), except $x$, $z$, and $v$, from Decaux et al. (1997).
[b] Oscillator strength $f_{lu}$, from Palmeri et al. (2003).
[c] Theoretical fluorescence yield, from Palmeri et al. (2003).

approximately 6.9 binary orbits. Several features of this spectrum are noteworthy; note that we indicate the rest-frame energies for all features in Figure 1, not yet taking account of orbital Doppler shifts. We see strong $n = 1–2$ line emission from the He- and H-like ionization stages of Fe. The H-like Fe Ly$\alpha$ transition is marginally resolved into its two fine-structure components. The four prominent $n = 1–2$ transitions in He-like Fe are resolved: the resonance line ($w$), the two intercombination lines ($x$, $y$), and the forbidden line ($z$) (Gabriel & Jordan 1969; Porquet et al. 2010). The resonance line shows a prominent P Cygni profile. We have also marked the rest-frame positions of the inner-shell transitions in the lower-charge states, close to the He-like $n = 1–2$ spectrum. These are the lines marked $v$, $u$, $r$, $q$, $t$, and $\beta$. We list all relevant transitions in the He-, Li-, and Be-like ions in the photon energy range 6600–6700 eV in Table 2.

As expected, the general character of the discrete Fe spectrum is very clearly that of a radiation-driven plasma (Kawashima & Kitamoto 1996; Liedahl & Paerels 1996; Paerels et al. 2000; Kallman et al. 2019). To date, the only detailed astrophysical Fe K emission spectra observed have been those of the Sun (solar flares, e.g., Phillips 2004) and of the core of the Perseus cluster, which was observed with the microcalorimeter spectrometer on Hitomi in 2016 (Hitomi Collaboration 2016). Both of these are examples of plasmas in which collisional excitation and ionization dominate.

In a collisional plasma, collisional excitation dominates over recombination-driven excitation, and the resonance line, $w$, is brighter than the forbidden and intercombination lines. But the latter are prominent in the spectrum of Cyg X-3, as is expected in a photoionized plasma in which recombination excitation dominates (Porquet et al. 2010). In pure photoionization/recombination equilibrium in low-density plasma, the forbidden line is expected to be much brighter than the intercombination lines. In their analysis of the spectrum, Kallman et al. (2019) converged on a model for the photoionized plasma that predicts the Fe XXV $n = 1–2$ emission-line intensity ratios $(x + y + z)/w = 12.1$ and $z/(x + y) = 1.7$; this model assumes that the plasma has low density and is optically thin. In Figure 3, we show the pure optically thin recombination spectrum that corresponds to the parameters derived by Kallman et al. (2019). From Figure 2, it appears that, instead, the intercombination lines are the strongest. The fact that the resonance line is observed to be much brighter than the photoionization model predicts suggests that the resonance line intensity is mostly produced by resonance scattering of continuum photons in the wind. The resonance line is in fact seen to exhibit a clear P Cygni profile.

The apparent observed "inverted" ratio of forbidden to intercombination line emission suggests that, as is the case in other sources (see Porquet et al. 2010 and references therein), the population of the upper state of the $z$ transition may be transferred to the upper states of the $x$, $y$ transitions faster than spontaneous radiative decay in the $z$ transition can occur. This can be brought about by electron thermal collisions at high densities or by the absorption of photons. The energy difference between the upper states of $z$ and $x$, $y$ in Fe XXV is 45.7 and 31.0 eV (i.e., the extreme ultraviolet or EUV), respectively, and it therefore requires a plasma of temperature $kT_e \sim$ a few dozen electronvolts, or an intense EUV radiation field, for this process to operate. Alternatively, the inner-shell transitions $n = 1–2$ in the Li- and Be-like ions may scatter continuum photons in the wind, giving rise to P Cygni line profiles, as is seen in the resonance line. We examine each mechanism in turn.

We first examine transfer between the upper states of $z$ and $x$, $y$ by photoexcitation by EUV photons. Such photons could originate in the accretion flow or in the photosphere of the W-R star. The spontaneous decay rate for the $1s2s\ ^3S_1$ level to ground is $A_{21} = 2.0 \times 10^8$ s$^{-1}$ (Drake 1971). The $1s2s\ ^3S_1 - 1s2p\ ^3P_{0,1,2}$ photoexcitation rate in a radiation field with energy flux density $F_\nu$ (erg cm$^{-2}$ s$^{-1}$ Hz$^{-1}$) is (rate, per atom per unit time):

$$R_{PE} = \frac{\pi e^2}{m_e c} f \frac{F_\nu}{h\nu}, \quad (1)$$

with $\nu$ being the frequency of the transition, $h$ Planck's constant, $f$ the oscillator strength, $e$ the electron charge (electrostatic unit), $m_e$ the electron mass, and $c$ the velocity of light. The oscillator strength for the $1s2s\ ^3S - 1s2p\ ^3P$ transition is $f = 0.032$ (Sanders & Kight 1989). Assuming an accretion luminosity $L$, emitted approximately as a blackbody with $kT_{BB} = 2$ keV (Kallman et al. 2019), we find a radiating surface area $A_{BB} = 7.3 \times 10^{12} L_{38}$ cm$^2$, with $L_{38}$ being the luminosity in units $10^{38}$ erg s$^{-1}$. At $\nu = 9.1 \times 10^{15}$ Hz, we find a photon flux at distance $r$ from the source $F_\nu/h\nu = 1.8 \times 10^{25} L_{38}/r^2$ photons cm$^{-2}$ s$^{-1}$ Hz$^{-1}$. In the Fe XXV ionization zone, we expect a photoionization parameter $\xi \equiv L/(nr^2) \approx 10^4$ erg cm s$^{-1}$ (Kallman & Bautista 2001). Substituting, we find $F_\nu/h\nu = 1.8 \times 10^4 n_{13} \xi_4$, and thus $R_{\rm PE} = 15.3 n_{13} \xi_4$ s$^{-1}$. Here, $\xi_4 \equiv \xi/10^4$, and we have used a fiducial value for the density in the W-R wind near the compact object of $n = 3 \times 10^{13} \dot{M}_{-5} v_{1000}^{-1} r_{11}^{-2}$ cm$^{-3}$, with $\dot{M}_{-5}$ being the mass-loss rate from the W-R star in $10^{-5} M_\odot$ yr$^{-1}$, $r_{11}$ being the distance from the center of the W-R star in units $10^{11}$ cm, and $v_{1000}$ being the wind velocity in units 1000 km s$^{-1}$; this estimate is based on a simple constant-velocity outflow. We conclude that the photoexcitation rate due to accretion radiation is much too small to suppress the forbidden line. For an analogous calculation for the rate induced by photospheric radiation from the W-R star, we assume $T_{\rm eff} = 100,000$ K, $\log L/L_\odot = 5.3$, and use model MW/WNL H20 13–19 from Todt et al. (2015), quoted in Koljonen & Maccarone (2017), for the W-R photospheric emission spectrum. These are likely to be upper limits to the stellar effective temperature and luminosity. We find for the photoexcitation rate $R_{\rm PE} = 4.7 \times 10^5 (F_{*,\lambda}/10^{-5}) r_{11}^{-2}$ s$^{-1}$, where $F_{*,\lambda}/10^{-5}$ is the W-R photospheric flux near 38 eV (326 Å) in units $10^{-5}$ erg cm$^{-2}$ s$^{-1}$ Å$^{-1}$. We conclude again that photoexcitation is not sufficient to suppress z; note that this rate cannot be increased much, since the stellar radius of the W-R star is in fact of order $10^{11}$ cm, and it appears unlikely that the photospheric EUV flux has been underestimated by a factor of 200 or more.

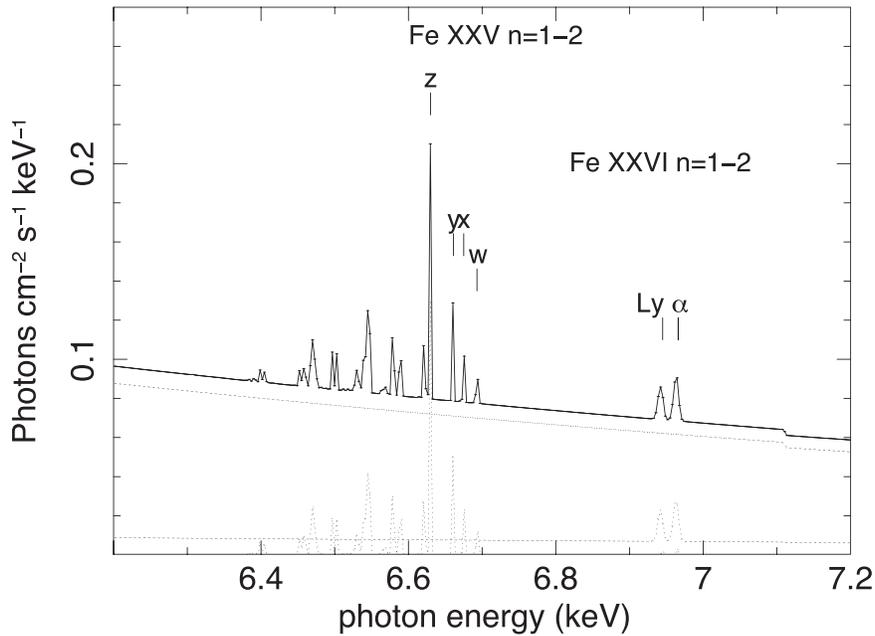

**Figure 3.** Emission from an optically thin pure recombining plasma, calculated from a model that describes most of the spectrum of Cyg X-3 at lower energies (Kallman et al. 2019). As seen in the He-like ions of the light- and mid-Z elements, the forbidden transition (z) in He-like Fe should be by far the brightest, with the resonance line w the weakest.

We now examine collisional transfer by thermal electrons in the photoionized gas. The temperature in the Fe XXV zone is expected to be of order 100 eV (Kallman & Bautista 2001), so electrons with sufficient kinetic energy are available. The density required for electron collisions to compete with spontaneous deexcitation of the $1s2s\,^3S_1$ state in the He-like Fe ion is approximately $n_e \sim 10^{17}$ cm$^{-3}$ (Porquet et al. 2010). The fiducial density of the wind near the compact object is of order $10^{13}$ cm$^{-3}$, however, so if collisional transfer is indeed responsible for the observed low value of the ratio of the forbidden to recombination line intensities, the emission would have to arise from very dense clumps very close to the compact object, so that one can have $\xi \sim 10^4$ while $n \sim 10^{17}$ cm$^{-3}$. Such a large density contrast of order $10^3$–$10^4$ appears unreasonable.

The inner-shell transitions $n = 1$–2 in the Li- and Be-like ions have large oscillator strengths; in these highly charged ions, the presence of one or a few electrons in $n = 2$ does not change the $n = 1$–2 transition probability much. For the highest charge states in the L-shell ions, the fluorescence yield is high, and autoionization following excitation is unimportant. These highest-charge ions can therefore resonantly scatter continuum photons almost as effectively as w. We suggest that this effect is what we observe in the spectrum of Cyg X-3: emission in the spectral range occupied by the intercombination lines and the resonance line is dominated by resonance-scattering-driven emission in the wind of the W-R star, by the w line and the inner-shell transitions in the Li- and possibly Be-like ions. Only (part of) the spectral window, near the forbidden transition, is not affected by radiative transfer.

At slightly lower energies, inner-shell absorption by B-like and lower charge states will often result in the destruction of the photons, because the ion relaxes by autoionization ("resonant Auger destruction"; Ross et al. 1996), and this may explain the shallow apparent absorption trough between 6450 and 6600 eV. The corresponding ranges of $n = 1$–2 transitions in the B- through F-like ions observed in laboratory experiments have been indicated in Figure 2 (Decaux et al. 1997; their Table 1). Kallman et al. (2019) pointed out the presence of this absorption in their analysis of the first-order HETGS spectrum. We also note that net absorption may also result if the absorbing material is not spherically symmetrically distributed, even if the photons are only scattered and not destroyed.

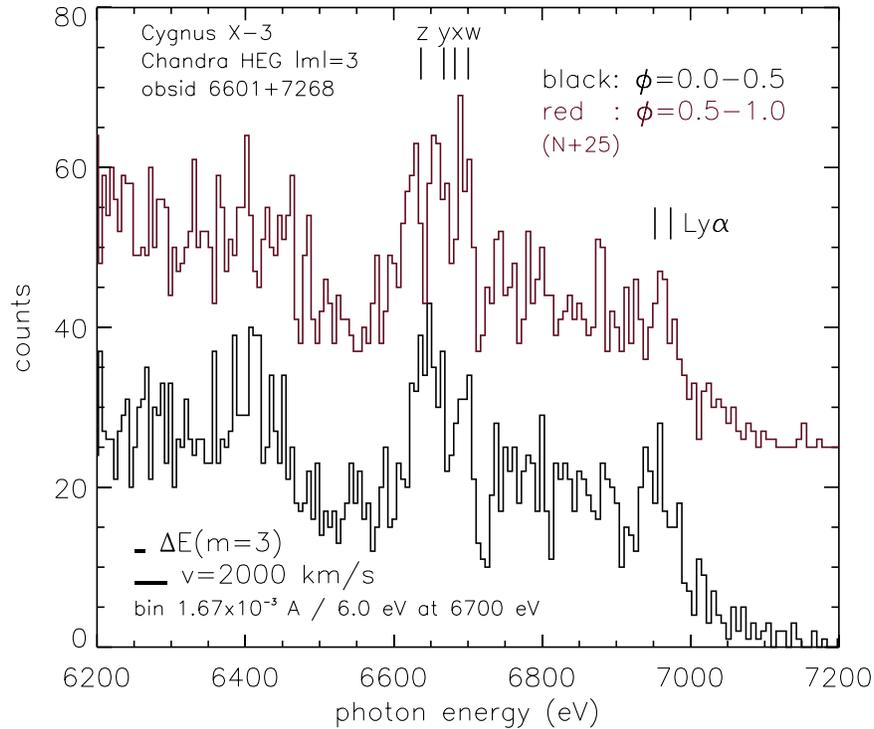

**Figure 4.** The same as Figure 2, but separated into two orbital phase bins and binned in twice wider wavelength bins. Phase 0.0–0.5 is when the compact object is approaching and phase 0.5–1.0 is when it is receding. The red phase 0.5–1.0 spectrum has been offset by 25 counts per bin, for clarity.

Finally, we note that the radiative recombination continua that accompany the discrete line spectrum have considerable diagnostic value (see, for instance, Paerels et al. 2000). However, in the case of the He- and H-like Fe spectra, these continua occur at such high energies (8.828 and 9.278 keV, respectively) that the signal-to-noise ratio is too poor to detect them in the present data.

## 4. Doppler Modulation of the He- and H-like Fe Line Complexes

The Fe XXV and Fe XXVI emission lines should arise closest to the compact object. In photoionization equilibrium, the He-like ionization zone is expected to be located at a distance $r \sim 3 \times 10^{10} L_{38}^{1/2} n_{13}^{-1/2} \xi_4^{-1/2}$ cm, while the binary separation is $a = 2.7 \times 10^{11}$ cm, if we assume equal masses of $10 M_\odot$ for both binary components. Note that for recombination-driven emission by the ionization stage $i$, the source for the emission is the $i+1$ ionization stage, which places the emitting plasma closer to the source than the nominal ionization zone for stage $i$. The emission lines from the highest ionization stages of Fe therefore in principle offer the best way to constrain the orbital motion of the compact object, and hence its mass. In the previous section, however, we have found that the emission in the lines with the largest transition probabilities is likely to be affected by the scattering of continuum photons and the subsequent resonance scattering. The observed emission profile of the $w$ line shows direct evidence for this effect, by exhibiting a P Cygni profile. The scattering of continuum photons will still produce emission coupled to the distribution of the scattering ions, but multiple scatterings may couple different regions of the wind, with a likely complicated distribution with respect to the center of mass of the binary and a complex orbital velocity structure. This argument disqualifies the $w$ line from consideration, and likely also the Fe XXVI Ly$\alpha$ lines.

Our original motivation for detailed analysis of the He-like spectrum was the fact that the intercombination and forbidden lines have much smaller transition probabilities, and are therefore most likely to remain optically thin, and should trace the motion of the compact object. But we find that the presence of emission from the Li- and Be-like ions complicates this idea: the spectral range of the $x$, $y$, and $z$ lines is likely contaminated by scattered light in inner-shell transitions of the Li- and Be-like ions. In the absence of a detailed model for the full distribution of the ions around the X-ray source, possibly augmented by a detailed radiative-transfer model, we cannot quantitatively account for this contamination of the He-like $x$, $y$, and $z$ emission spectra.

But since there is no other probe of the motion of the compact object, we will still formally analyze the phase-resolved spectrum for Doppler shifts. A full model for the excitation and radiative transfer through the wind would be needed to interpret the measurements in terms of the properties of the compact object. But if the scattering in the Li- and Be-like ions is modest, and the Li- and Be-like ionization zones are approximately spherically symmetric and centered on the compact object, any Doppler shift could be interpreted more directly. We will therefore carry out a Doppler shift measurement for the spectral range of the $x$, $y$, and $z$ lines.

In view of the limited statistics of the spectra, we chose to cross-correlate the spectrum binned in two phase intervals covering the expected largest red- and blueshift of the Fe lines (which should occur at phases 0.75 and 0.25, respectively).

In Figure 4, we display the high-resolution spectrum separated into two orbital phase bins: phase $0.0 - 0.5$, when the compact object is approaching us (the compact object is in front at phase 0.5), and $0.5 - 1.0$ when the compact object is receding.

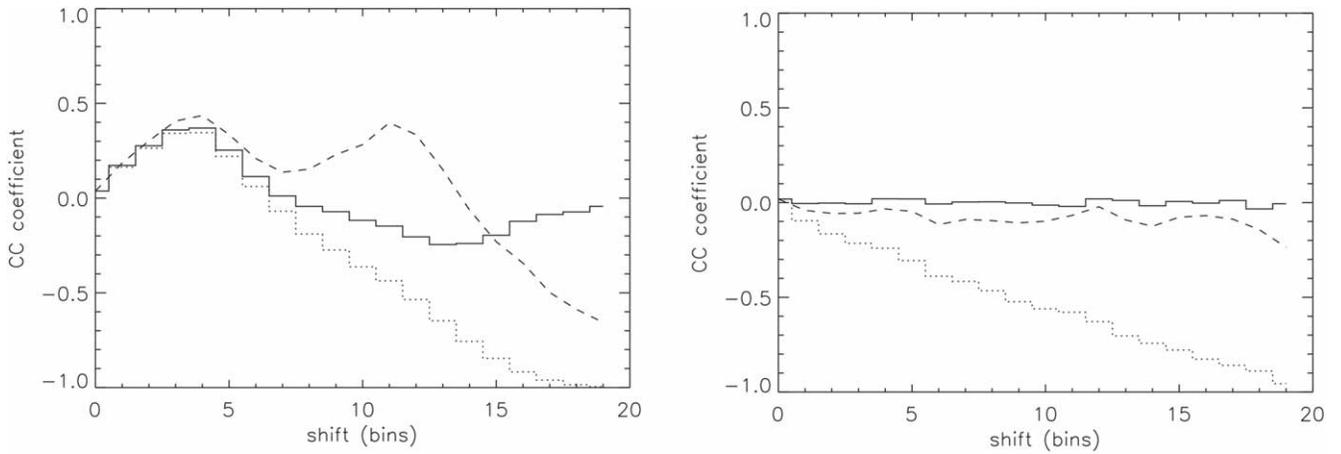

**Figure 5.** Left: cross-correlation coefficient between two synthetic emission-line spectra, as a function of Doppler shift. The solid line is the "circular" cross-correlation, the dotted line is the cross-correlation with the shifted histogram padded out with zeros, and the dashed line is the same, but normalized only over nonzero bins. The right-hand panels shows the same, but for a flat background only.

We computed the cross-correlation between the two spectra, as a function of wavelength shift, wrapping the shifted spectrum around, instead of padding out the spectra with zeros beyond the range of the He-like spectrum. This ensures a constant expectation value for the correlation coefficient for pure noise, instead of increasing noise fluctuations with increasing shift (decreasing overlap).

We first examine the properties of this cross-correlation algorithm. The cross-correlation coefficient $C_j$ between two quantities $f_i$ and $g_{i+j}$ (both in units of counts bin$^{-1}$) is calculated as

$$C_j = N_{\rm bins}\frac{\sum_i f_i \; g_{i+j}}{\sum f_i \cdot \sum g_i} - 1. \quad (2)$$

We made a synthetic spectrum $f_i$ consisting of three Gaussian emission lines at the energies of Fe XXV $w$, $x$, and $y$, with width $\sigma = 8$ eV. The energy range spanned 6600–6720 eV, in 6 eV bins. Each emission line contained 150 counts. We then applied a shift of $\Delta v = \pm 500$ km s$^{-1}$ to two copies of the set of emission lines. We added a constant background of 15 counts bin$^{-1}$, and we then drew from a Poisson distribution for each bin in the resulting histogram. The cross-correlation was computed as in Equation (2), in three different ways: by wrapping around the shifted histogram ("circular"), by padding the shifted histogram with zeros ("padded"), and by padding the shifted histogram but normalizing only over nonzero bins ("padded-prime"). We then repeated this procedure for a pure noise background, at an expectation value of 15 counts bin$^{-1}$. We display the results in Figure 5. The cross-correlation peaks at a shift of 3.7 bins, or 1000 km $^{-1}$. As can be seen, the "circular" cross-correlation has the desirable properties of being flat in the absence of correlation and avoiding the division by small numbers that affects a cross-correlation using the histogram padded out with zeros and subsequently renormalized over nonzero bins.

In order to estimate the Doppler modulation of a group of lines or any individual line, we actually fit a Gaussian to the line in both phase-binned spectra and proceeded to evaluate the circular cross-correlation of the two phase-binned Gaussian fits. We opted for this route instead of directly cross-correlating the data, as both the Fe XXVI Ly$\alpha$ and Fe XXV He$\alpha$ spectra are very noisy in the third order. This has the same effect as smoothing the cross-correlation as a function of shift, with a smoothing width of order of the width of the lines. The discrete circular cross-correlation was evaluated by wrapping the two spectra onto each other and then evaluating the discrete cross-correlation. The shift at which the cross-correlation coefficient reaches a statistically significant ($>3\sigma$) global maximum is considered to be the Doppler shift between the two spectra. In order to avoid skewing the standard deviation ($\sigma$) by a huge margin, the values of the bins that have not been acted upon by the Levenberg–Marquardt Gaussian fitting algorithm have been set to a value greater than or equal to the average photon count in those regions. The expected variance in the cross-correlation coefficient was estimated by propagating Poissonian counting statistics.

In Figure 6, we show the Gaussian fits to Fe XXVI Ly$\alpha_1$ and Ly$\alpha_2$, for the two phase bins. Figure 7 shows the cross-correlation coefficient between these two fits, as a function of wavelength shift. We calculated the cross-correlation coefficient over the 6945–6990 eV range, and shifted in increments of 3 eV, or $\Delta v = 128$ km s$^{-1}$. From Figure 7, we can conclude that the cross-correlation coefficient for the two Fe XXVI Ly$\alpha$ lines reaches its global maximum (statistically significant) at zero Doppler shift. The cross-correlation falls below the $3\sigma$ deviation from zero (no correlation) at a shift of $\approx$1.5 velocity bins or $\Delta v = 192$ km s$^{-1}$ (the full offset between one spectrum and the other, so double the orbital velocity of the emission-line source). Vilhu et al. (2009) determined an orbital velocity from the Ly$\alpha$ lines, using the first-order HEG spectrum, of $418 \pm 123$ km s$^1$. For both the first- and third-spectral-order radial velocity measurements in Ly$\alpha_{1,2}$ to be consistent, we would have to assume that the radial velocity amplitude is in fact near the upper end of our upper limit.

Figure 8 shows the corresponding Gaussian fits to the Fe XXV spectrum (fits to the intercombination and forbidden lines). In Figure 9, the cross-correlation coefficient for the $x$, $z$ lines reaches a statistically significant global maximum (=0.78) at a shift of $\approx$2 bins or $\Delta v = 270$ km s$^{-1}$. The full range of shifts consistent with positive correlation at more than $3\sigma$ is 0 −4 bins, or 0–540 km s$^{-1}$. There also appears to be a significant apparent anticorrelation at shifts of 8 and 25 bins, or 1080 and 3375 km s$^{-1}$, respectively. This probably reflects the fact that there is in fact significant emission from scattering by the inner-shell transitions in the Li- and Be-like ions, which are likely to have a nontrivial orbital phase dependence.

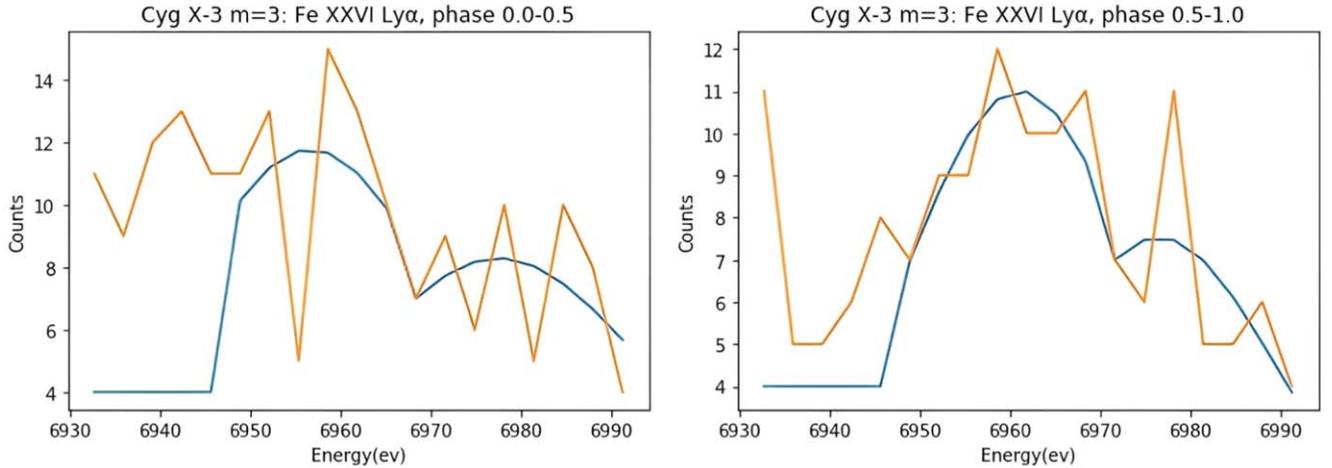

**Figure 6.** Gaussian fits (blue) for the Lyman α1 line (∼6.95 keV) and the Lyman α2 line (∼6.97 keV), superimposed on the respective spectra (orange): phase 0.0–0.5 (left) and phase 0.5–1.0 (right).

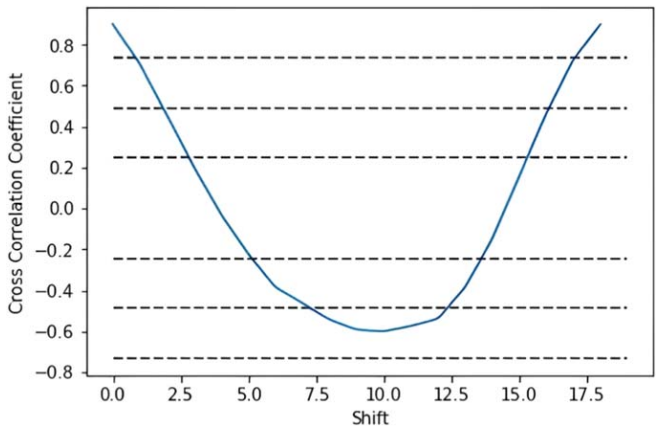

**Figure 7.** Cross-correlation plot of the two plots in Figure 5. On the vertical axis, we plot the normalized cross-correlation between the two spectra. The six dotted lines above and below zero represent values of the cross-correlation ±1σ, 2σ, and 3σ away from the expectation value for pure noise. The shift is in units of 3 eV, or $\Delta v \cong 128$ km s$^{-1}$ at 7000 eV.

We find that the Fe XXVI Lyα lines, integrated over the phase intervals 0.0–0.5 and 0.5–1.0 (compact object approaching and receding, respectively), appear to show no significant Doppler offset between the two phase intervals, with the maximum positive correlation between the spectra occurring for a range of velocity shifts of 0−192 km s$^{-1}$ (3σ). We will try to estimate the effect of integrating over such a wide phase interval, by computing the average of the radial velocity over half an orbit, in units of the maximum radial velocity. For uniform emissivity as a function of phase, that reduction is by a factor $2/\pi$, so that our estimate for the maximum allowed radial velocity range for the compact object is $\pi/2 \times 192 = 300$ km s$^{-1}$, or a maximum orbital velocity (assuming a circular orbit) of $v_{\rm orb} = 150$ km s$^{-1}$.

The same reasoning applied to the intercombination and forbidden emission lines in the He-like ion yields a maximum orbital velocity range of 848 km s$^{-1}$, or a maximum orbital velocity of $v_{\rm orb} = 424$ km s$^{-1}$.

If we take these limits at face value, as upper limits to the orbital velocity of the compact object, we arrive at the following conclusion. The fact that the compact object does not exhibit a measurable radial velocity in our data translates into a lower limit on its mass, $M_C$, of $M_C \geqslant (v_{\rm rad,W-R}/v_{\rm rad,C}) M_{\rm W-R}$, with $v_{\rm rad,W-R}$ being the orbital velocity of the W-R star, $M_{\rm W-R}$ its mass, and $v_{\rm rad,C}$ our upper limit to the orbital velocity of the compact object. Koljonen & Maccarone (2017) recommend a range for the probable mass of the W-R star of ($M_{\rm W-R} = 8$–$15 M_\odot$), while the radial velocity (half the full range) is $K = 379^{+124}_{-149}$ km s$^{-1}$. That translates into a nominal lower limit on the mass of the compact object of $M_C = 20$–$38 M_\odot$ using the Lyα lines, or $M_C = 7.2 - 13.4 M_\odot$ using the He-like Fe forbidden and intercombination lines. We caution again that the spectrum has significant blending with the emission from the Li- and Be-like ions, and this systematic error likely dominates over any random noise. We note that higher spectral resolution may not straightforwardly alleviate this problem, due to the large velocity broadening of the emission lines.

## 5. Discussion

We extend the spectroscopic study of the X-ray line emission of Cyg X-3 from the mid-Z elements to the Fe K band, using the third diffraction order of the Chandra HETGS/HEG spectrum. Previous studies (Paerels et al. 2000; Kallman et al. 2019) have shown that the discrete emission in the mid-Z elements arises from recombination excitation in the photoionized wind of the W-R star.

The spectrum in the 6000–7500 eV band shows $n = 1$–2 line emission from the He- and H-like Fe ions. The intensity ratios of the four $n = 1$–2 transitions in the He-like ion suggest that there is a contribution from inner-shell emission from the Li- and possibly the Be-like ions as well. We detect a clear P Cygni profile in the He-like $n = 1$–2 resonance line ($w$), direct evidence confirming that the line emission is excited in the highly ionized W-R wind, in the vicinity of the compact object. We detect broad absorption in the 6400–6600 eV band, which may be due to multiple inner-shell transitions in the lower-charge members of the Fe L shell series.

The appearance of a P Cygni profile in the He-like resonance line indicates that there is likely to be a finite contribution in emission from the resonance scattering of continuum photons in the wind. By extension, a similar effect may contribute to the H-like Ly$\alpha_{1,2}$ emission lines (we do not have sufficient flux to detect a P Cygni signature in these lines, directly probing the optical depth of the wind). The only optically thin transitions likely to trace the motion of the compact object are therefore the He-like forbidden transition (*z*) and possibly the intercombination lines (*x*, *y*).

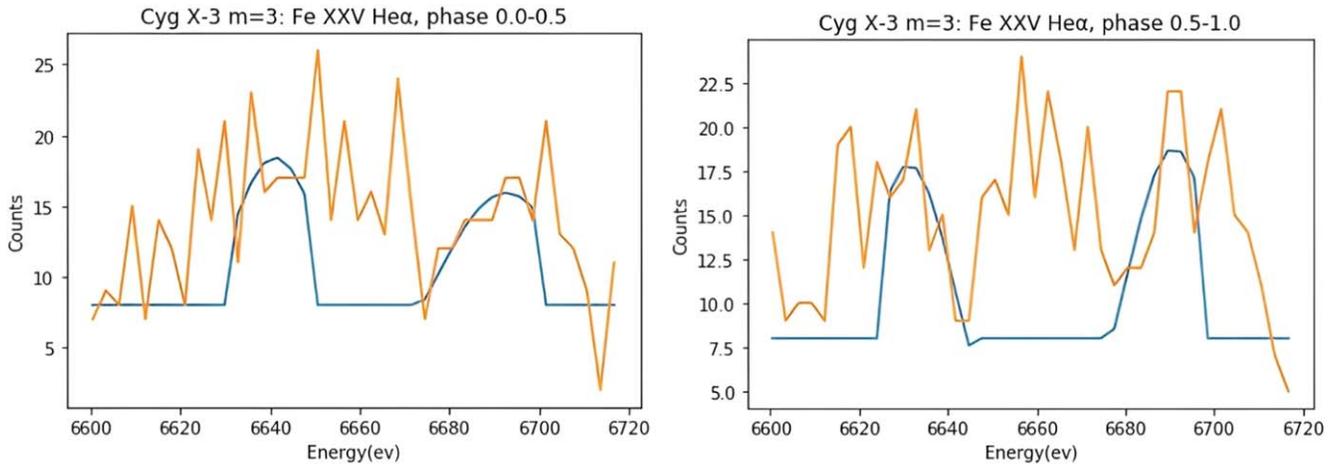

**Figure 8.** Gaussian fits (blue) for the *x* intercombination line (∼6.68 keV) and the *z* forbidden line (∼6.63 keV), superimposed on the respective spectra (orange): phase 0.0–0.5 (left) and phase 0.5–1.0 (right).

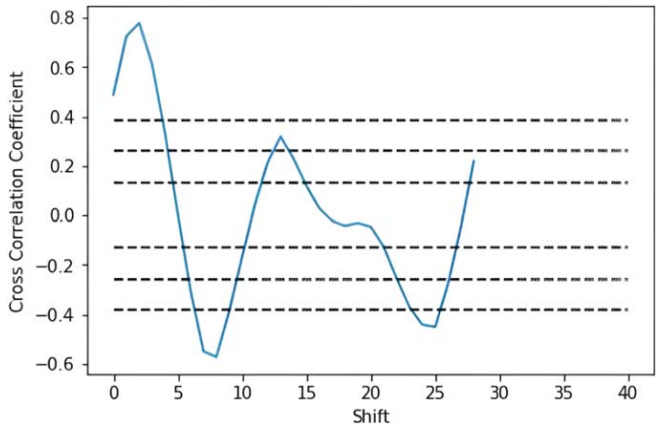

**Figure 9.** Cross-correlation plot of the two plots in Figure 6. The shift bin size is 3 eV, or $\Delta v = 135$ km s$^{-1}$ at 6650 eV. The six dotted lines above and below 0.0 represent values of the cross-correlation $\pm 1\sigma$, $2\sigma$, and $3\sigma$ away from the expectation value for pure noise.

The intensity ratios in the He-like $n = 1$–2 transitions do not conform to the expectation for an optically thin, low-density recombining plasma. The *w* line, as we noted above, clearly shows a contribution from the resonance scattering of continuum photons, so it is much brighter than expected from pure recombination. The forbidden line would be expected to be the brightest of the four $n = 1$–2 transitions. Instead, we see comparable fluxes in *z*, *x*, and *y*. We show that the usual mechanisms that can transfer flux from *z* to *x* and *y* (radiative excitation by UV photons or collisional excitation by thermal electrons) are not viable in this case. This suggests there may be a finite contribution from resonance scattering by the inner-shell transitions of the Li- and Be-like ions, of the type $1s^22s - 1s2s2p$ and $1s^22s^2 - 1s2s^22p$, respectively, which have similar transition energies to the He-like forbidden and intercombination lines.

Though we suspect contamination by scattered photons (and noting that recombination-excited and scattered photons arise from different ionization stages), we performed a radial velocity study on the H-like Ly$\alpha$ lines. In the He-like spectrum, we used the range for the *x* and *z* lines (*x* and *z* have very small radiative transition probabilities). We cross-correlated the emission-line profiles obtained in two phase bins, centered on the maximum approaching and receding radial velocity of the compact object. We find no clearly detectable Doppler modulation. In fact, the Doppler shift between the two phases is consistent with zero, with upper limits on the orbital radial velocity of the compact object of 150 km s$^{-1}$ in Ly$\alpha_{1,2}$ (assuming a circular orbit). The corresponding upper limit to the orbital radial velocity of the compact object in the He-like lines is 424 km s$^{-1}$. With no radial velocity modulation detected, we then infer a lower limit on the mass of the compact object, simply from the ratio of the radial velocities of the binary components and an estimate for the mass of the W-R star, of $M_C \geqslant 7.2 - 38 M_\odot$, but bearing in mind that the emission-line spectrum may be complicated by radiative transfer.

### Acknowledgments

We gratefully acknowledge discussions with Tim Kallman, Duane Liedahl, and Mike McCollough. We wish to express our gratitude to our anonymous referee, whose careful critique of the paper helped to improve it significantly.

### ORCID iDs

Aswath Suryanarayanan  https://orcid.org/0000-0002-2167-3170

### References

Corrales, L. R., & Paerels, F. 2015, MNRAS, 453, 1121
Decaux, V., Beiersdorfer, P., Kahn, S. M., & Jacobs, V. L. 1997, ApJ, 482, 1076
Drake, G. W. 1971, PhRvA, 3, 908
Gabriel, A. H., & Jordan, C. 1969, MNRAS, 145, 241
Hatchett, S., & McCray, R. 1977, ApJ, 211, 552
Hitomi Collaboration 2016, Natur, 535, 117
Kallman, T., & Bautista, M. 2001, ApJS, 133, 221

Kallman, T., McCollough, M., Koljonen, K., et al. 2019, ApJ, 874, 51
Kawashima, K., & Kitamoto, S. 1996, PASJ, 48, L113
Koljonen, K. I. I., & Maccarone, T. J. 2017, MNRAS, 472, 2181
Liedahl, D., & Paerels, F. 1996, ApJL, 468, L33
Liu, Q. Z., van Paradijs, J., & van den Heuvel, E. P. J. 2006, A&A, 455, 1165
Paerels, F., Cottam, J., Sako, M., et al. 2000, ApJL, 533, L135
Palmeri, P., Mendoza, C., Kallman, T. R., & Bautista, M. A. 2003, A&A, 403, 1175
Phillips, K. J. H. 2004, ApJ, 605, 921
Porquet, D., Dubau, J., & Grosso, N. 2010, SSRv, 157, 103
Ross, R. R., Fabian, A. C., & Brandt, W. N. 1996, MNRAS, 278, 1082
Rudolph, J. K., Bernitt, S., Epp, S. W., et al. 2013, PhRvL, 111, 103002
Sanders, F. C., & Kight, R. E. 1989, PhRvA, 39, 4387
Stark, M., & Saia, M. 2003, ApJL, 587, L101
Tarter, C. B., Tucker, W. H., & Salpeter, E. E. 1969, ApJ, 156, 943
Todt, H., Sander, A., Hainich, R., et al. 2015, A&A, 579, A75
Vilhu, O., Hakala, P., Hannikainen, D. C., McCollough, M., & Koljonen, K. 2009, A&A, 501, 679